\documentclass{DISproc}

\usepackage{graphicx}
\usepackage{caption}
\usepackage{subcaption}

\begin{document}

\def\msbar{\overline{MS}}
\def\gsim{\mathrel{\rlap{\lower4pt\hbox{\hskip1pt$\sim$}} \raise1pt\hbox{$>$}}} 
\def\lsim{\mathrel{\rlap{\lower4pt\hbox{\hskip1pt$\sim$}} \raise1pt\hbox{$<$}}}  
\newcommand{\Ord}{\ensuremath{{\cal O}}}
\def\ZMVFNS{\rm ZMVFNS}


\title{
Heavy Quark Production in the ACOT Scheme Beyond NLO}

\author{
{\slshape
T.~Stavreva\rlap,${}^1$ 
F.~I.~Olness\rlap,${}^2$\footnote{Presented by F.I.~Olness
at the 
{\sl 20th International Workshop on Deep-Inelastic Scattering and Related Subjects (DIS 2012)},
26-30 Mar 2012: Bonn, Germany.}
 \null \  
I.~Schienbein\rlap,${}^1$ 
T.~Je\v{z}o\rlap,${}^1$ 
A.~Kusina\rlap,${}^2$ 
K.~Kova\v{r}\'{\i}k\rlap,${}^3$ 
J.~Y.~Yu${}^{1,2}$ 
}
\\[1ex]
${}^1$Laboratoire de Physique Subatomique et de Cosmologie, Universit\'e
Joseph Fourier/CNRS-IN2P3/INPG, 
 53 Avenue des Martyrs, 38026 Grenoble, France
\\
${}^2$Southern Methodist University, Dallas, TX 75275, USA
\\
${}^3$Institute for Theoretical Physics, Karlsruhe Institute of Technology, 
Karlsruhe, D-76128, Germany
}

\contribID{\null}

\doi  

\maketitle

\begin{abstract}
  We analyze the properties of the ACOT scheme for heavy quark production
and make use of the $\msbar$ massless results at NNLO and N$^{3}$LO
for the structure functions $F_{2}$ and $F_{L}$ in neutral current 
deep-inelastic scattering to estimate the higher order corrections. 
The dominant heavy quark mass effects at higher orders 
can be taken into account using the massless Wilson coefficients together with
an appropriate slow-rescaling prescription implementing the phase space constraints.
Combining the exact ACOT scheme at NLO with these expressions should provide
a  good approximation to the full calculation in the ACOT scheme 
at NNLO and N$^{3}$LO. 
\end{abstract}


\section{Introduction\label{sec:intro}}


With the ever-increasing precision of the experimental data, the
production of heavy quarks in high energy processes has become an
increasingly important subject.  
As theoretical calculations and
parton distribution function (PDF) evolution are progressing to
next-to-next-to-leading order (NNLO) of QCD, there is a clear need to
formulate and also implement the heavy quark schemes at this order and
beyond.
The most important case is arguably the heavy quark treatment
in inclusive deep-inelastic scattering (DIS) since the very precise
HERA data for DIS structure functions and cross sections form the backbone
of any modern global analysis of PDFs. Here, the heavy quarks
contribute up to 30\% or 40\% to the structure functions at small momentum
fractions $x$.
Extending the heavy quark schemes to higher orders is therefore 
necessary for extracting precise PDFs and hence for precise predictions
of observables at the LHC.
Additionally, it is theoretically important to have 
a general pQCD framework including heavy quarks which is valid to all orders in perturbation theory
over a wide range of hard energy scales.
The results of this study form the basis for using the ACOT scheme in NNLO global analyses
and for future comparisons with precision data for DIS structure functions.

\section{ACOT Scheme \label{sec:acot}}

The ACOT renormalization scheme \cite{Aivazis:1993pi} provides a mechanism
to incorporate the heavy quark mass into the theoretical calculation
of heavy quark production both kinematically and dynamically. In 1998
Collins \cite{Collins:1998rz} extended the factorization theorem to
address the case of heavy quarks; this work provided the theoretical
foundation that allows us to reliably compute heavy quark processes
throughout the full kinematic realm.

The ACOT
prescription is to just calculate the massive partonic cross sections
and perform the factorization using the quark mass as regulator.
The ACOT scheme  does not need any
observable--dependent extra contributions or any regulators to smooth
the transition between the high and low scale regions.


In Ref.~\cite{Stavreva:2012bs}
 we demonstrated using the  NLO  full ACOT scheme that the dominant mass effects 
are those coming from the phase space which can be  taken into account
via a generalized slow-rescaling $\chi(n)$-prescription.\footnote{Specifically, 
$\chi(n)=x[1+(n\, m/Q)^2]$, where $m$ is the quark mass.
See Ref.~\cite{Stavreva:2012bs} for details and a complete set of references. 
}
Assuming that a similar relation remains true at higher orders, one can construct
the following approximation to the ACOT result  up to N$^3$LO ($\Ord(\alpha_S^3)$):
\begin{eqnarray}
{\rm ACOT} [\Ord(\alpha_S^{0+1+2+3})] 
 &\simeq& 
 {\rm ACOT} [\Ord(\alpha_S^{0+1})] + \ZMVFNS_{\chi(n)} [\Ord(\alpha_S^{2+3})]
\label{eq:approx}
\end{eqnarray}
In this equation, 
``ACOT'' generically represents any variant of the 
ACOT scheme (ACOT, \hbox{S-ACOT}, \hbox{S-ACOT${}_\chi$});
for the results presented in Sec.~\ref{sec:numerics}, we will use the fully massive 
ACOT scheme with all masses retained out to NLO. 
The
$\ZMVFNS_{\chi(n)}$ term uses the massless Wilson coefficients at ${\cal O}(\alpha\,\alpha_{S}^{2})$
and ${\cal O}(\alpha\,\alpha_{S}^{3})$.
This approximation is necessary as 
not all the necessary massive Wilson coefficients at ${\cal O}(\alpha\,\alpha_{S}^{2})$
and ${\cal O}(\alpha\,\alpha_{S}^{3})$ have been computed.

\section{Results}
\label{sec:numerics}

%
\begin{figure*}[t]
\begin{subfigure}{1.0\textwidth}
\caption{$F_2^j/F_2$ vs. $Q$.}
\label{fig:f2RatX135-1}
\includegraphics[width=0.30\textwidth]{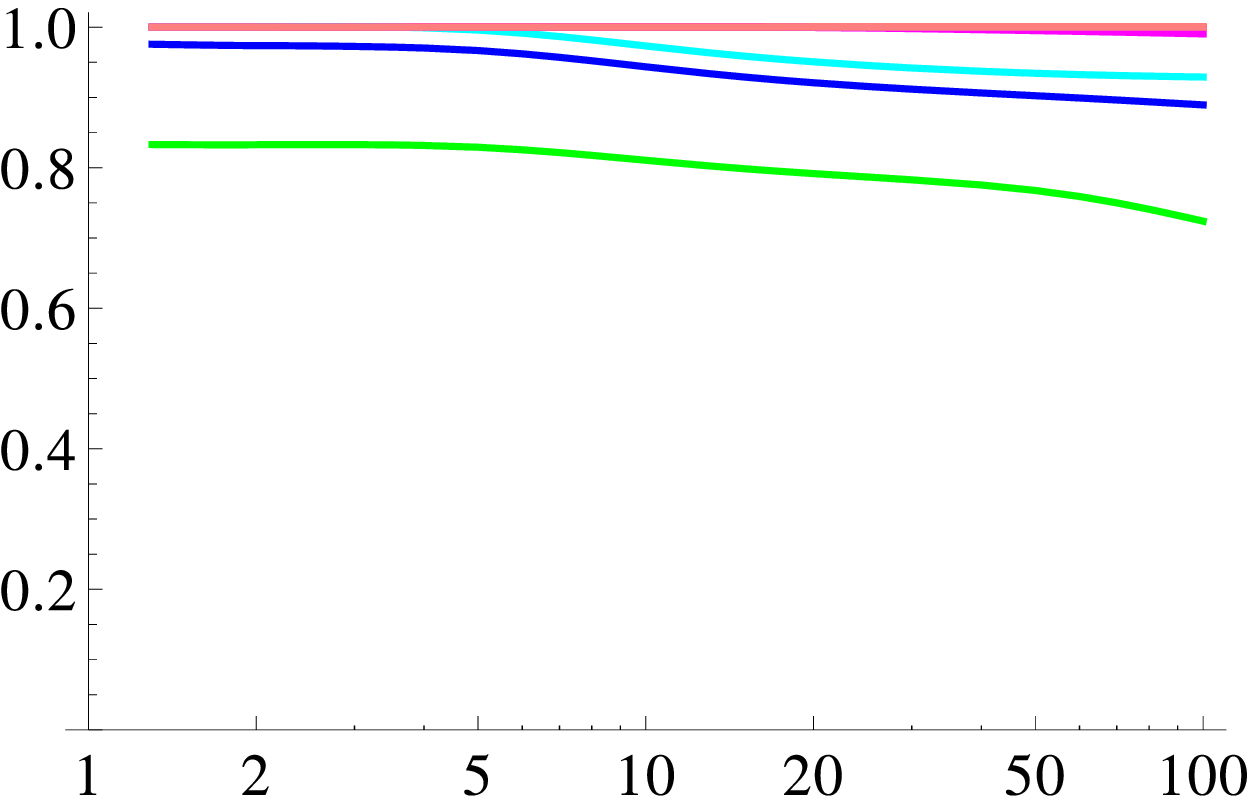}
\hfil
\includegraphics[width=0.30\textwidth]{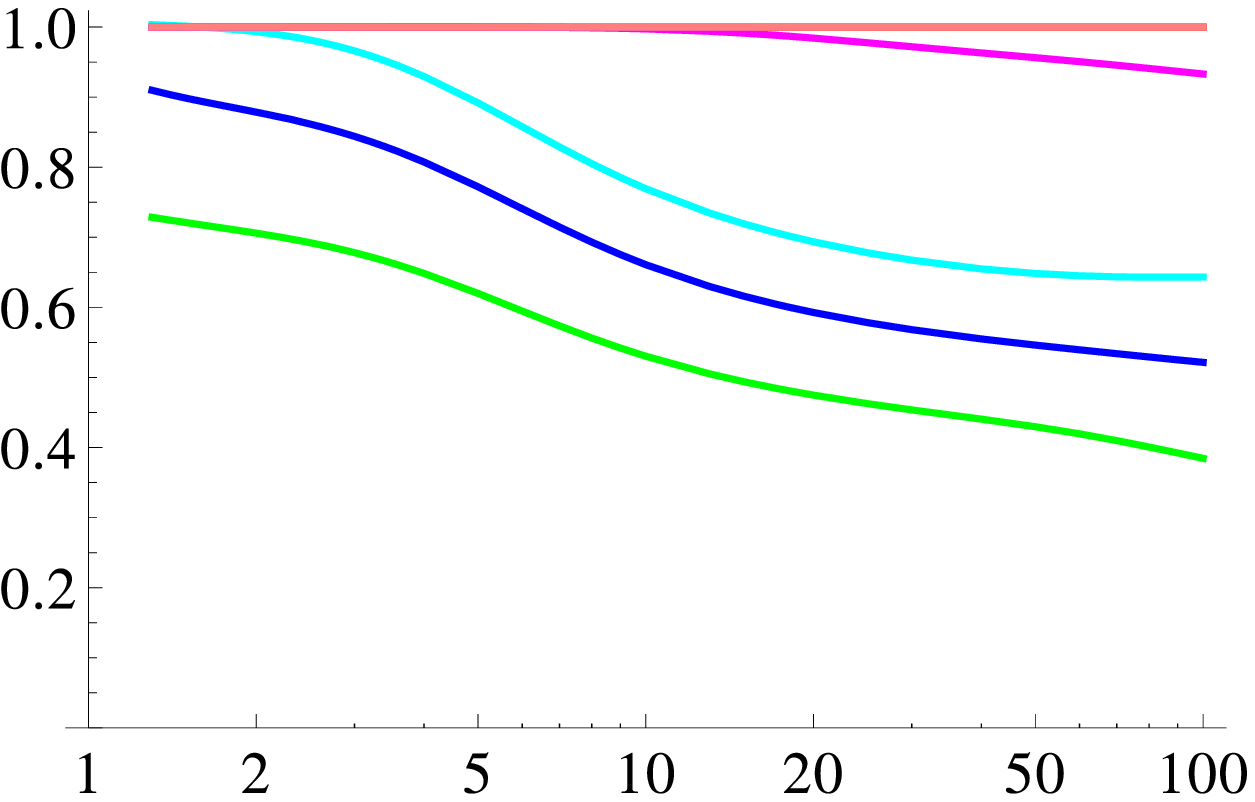}
\hfil
\includegraphics[width=0.30\textwidth]{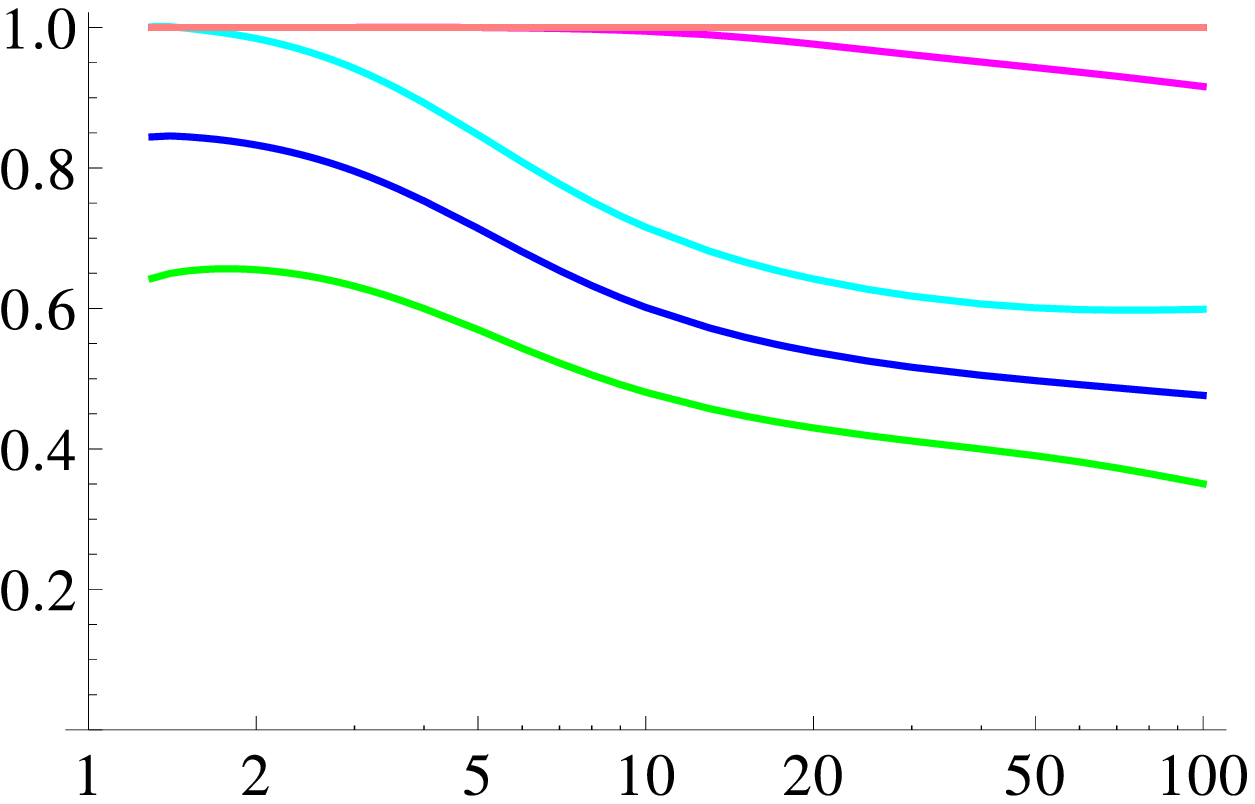}
\end{subfigure}
\begin{subfigure}{1.0\textwidth}
\caption{ $F_L^j/F_L$ vs. $Q$.}
\label{fig:fLRatX135-1}
\includegraphics[width=0.30\textwidth]{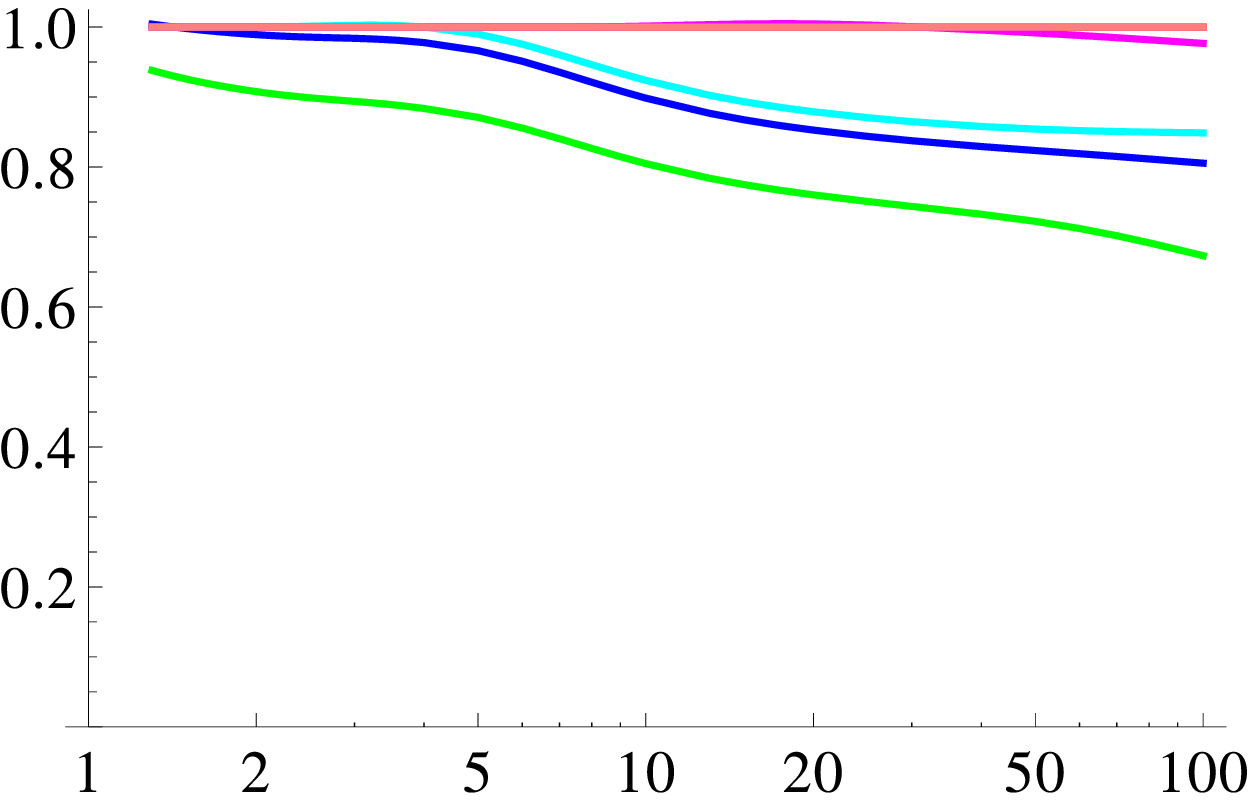}
\hfil
\includegraphics[width=0.30\textwidth]{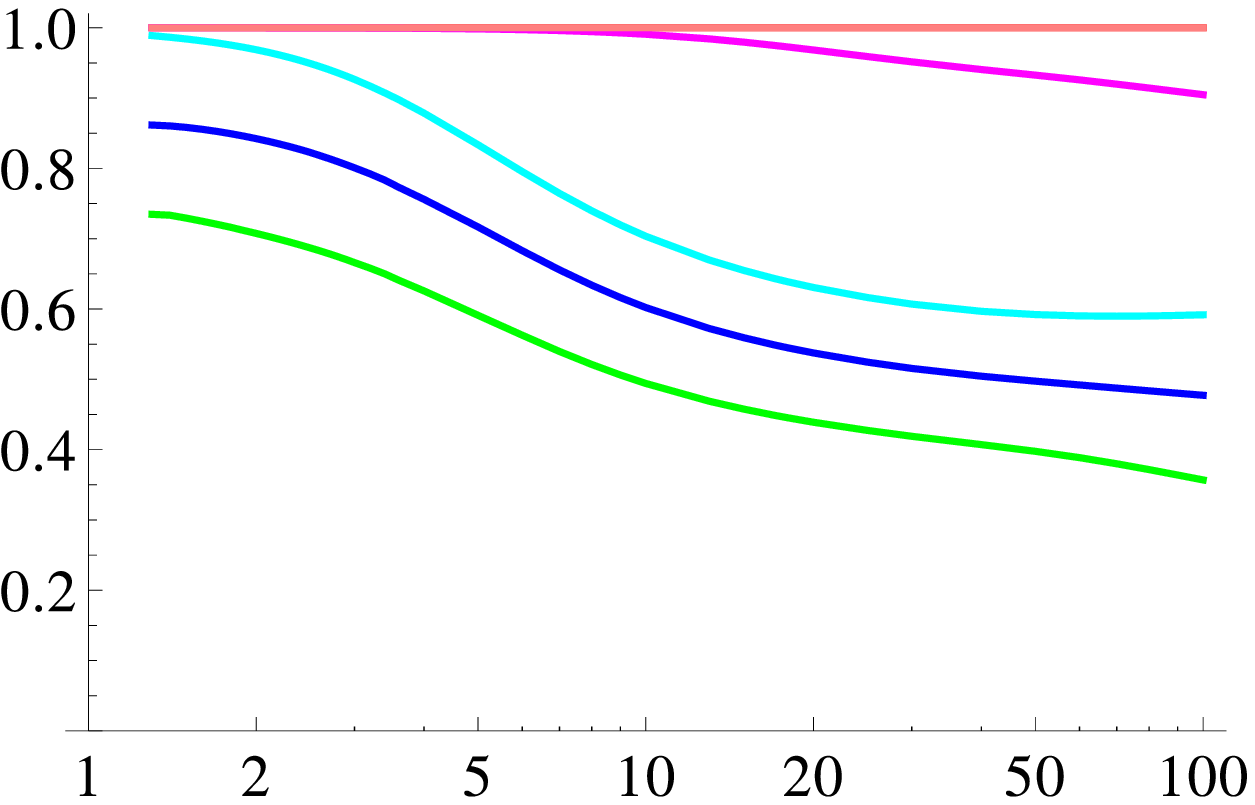}
\hfil
\includegraphics[width=0.30\textwidth]{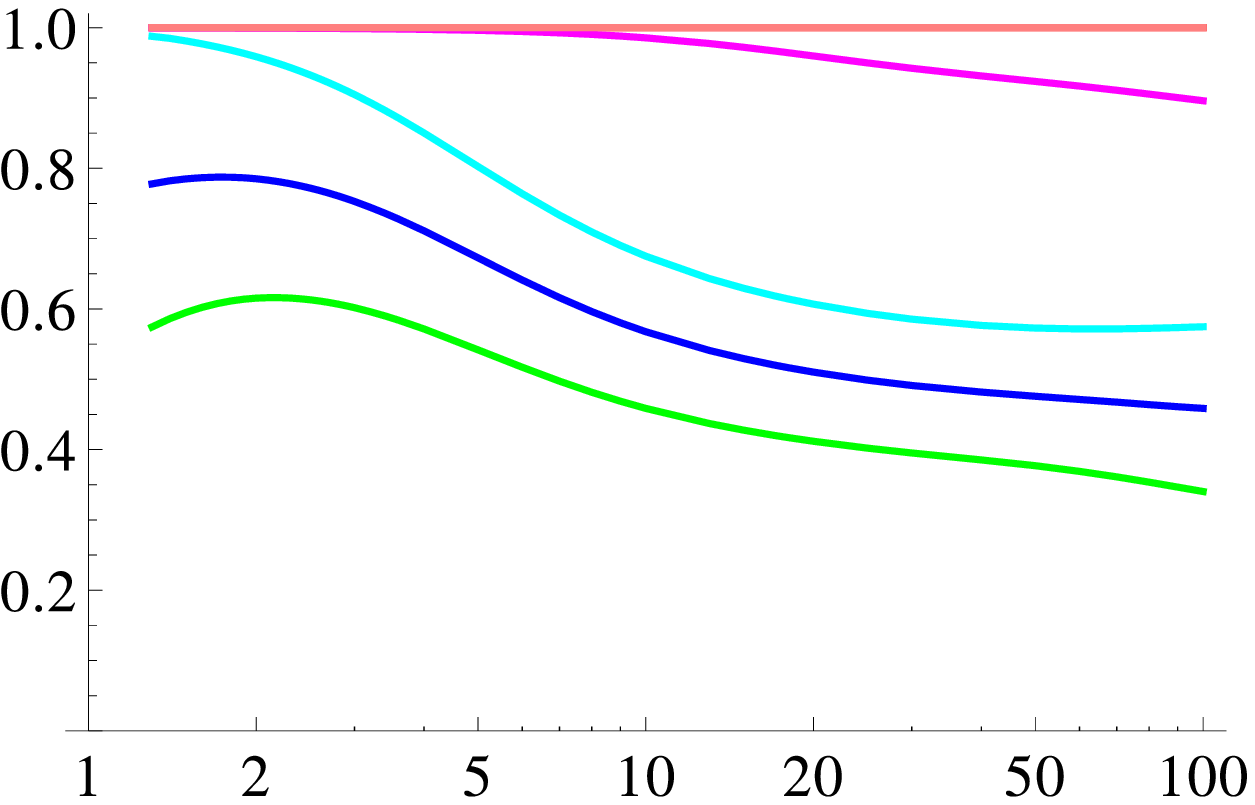}
\end{subfigure}
\caption{Fractional contribution for each quark flavor to
$F_{2,L}^j/F_{2,L}$ vs. $Q$ at N$^3$LO for fixed $x=\{10^{-1},10^{-3},10^{-5}\}$
(left to right). Results are displayed for $n=2$ scaling.
Reading from the bottom, we have the cumulative contributions from
the $j=\{u,d,s,c,b\}$ (green, blue, cyan, magenta, pink).}
\end{figure*}


We now present the results of our calculation extending the ACOT scheme to 
NNLO and N${}^3$LO. 
The details of the  numerical calculations are presented in Ref.~\cite{Stavreva:2012bs}.
We have used the  QCDNUM program \cite{Botje:2010ay} for the DGLAP evolution, 
and the Fortran subroutines provided by Andreas Vogt for the higher order Wilson coefficients. 
We choose $m_c=1.3$~GeV,  $m_b=4.5$~GeV,  $\alpha_S(M_Z)=0.118$.



In Figures \ref{fig:f2RatX135-1} and \ref{fig:fLRatX135-1} we display
the fractional contributions for the final-state quarks 
 to the structure functions $F_{2}$ and
$F_{L}$, respectively, for selected $x$ values as a function of
$Q$. 
We observe that for
large $x$ and low $Q$ the heavy flavor contributions are minimal. 
but these can grow quickly as we move to smaller $x$ and larger $Q$. 
For example,  at $x=10^{-5}$
and large $Q$ we see the $F_2$ contributions of the $u$-quark and $c$-quark
are comparable (as they both couple with a factor 4/9), and the $d$-quark
and $s$-quark are comparable (as they both couple with a factor 1/9).


%
\begin{figure*}[t]
\begin{subfigure}{1.0\textwidth}
\caption{$F_{2}$ vs. $Q$.}
\label{fig:f2orders}
\includegraphics[width=0.30\textwidth]{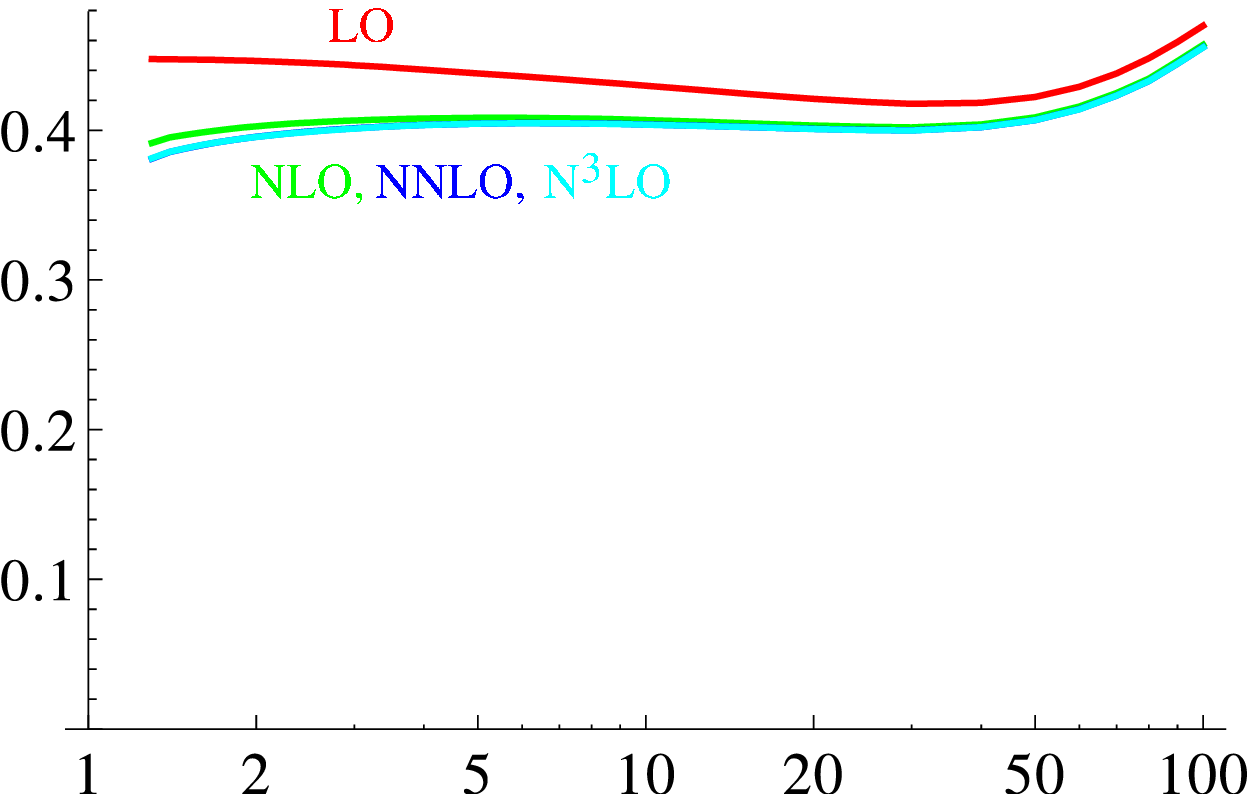}
\hfil
\includegraphics[width=0.30\textwidth]{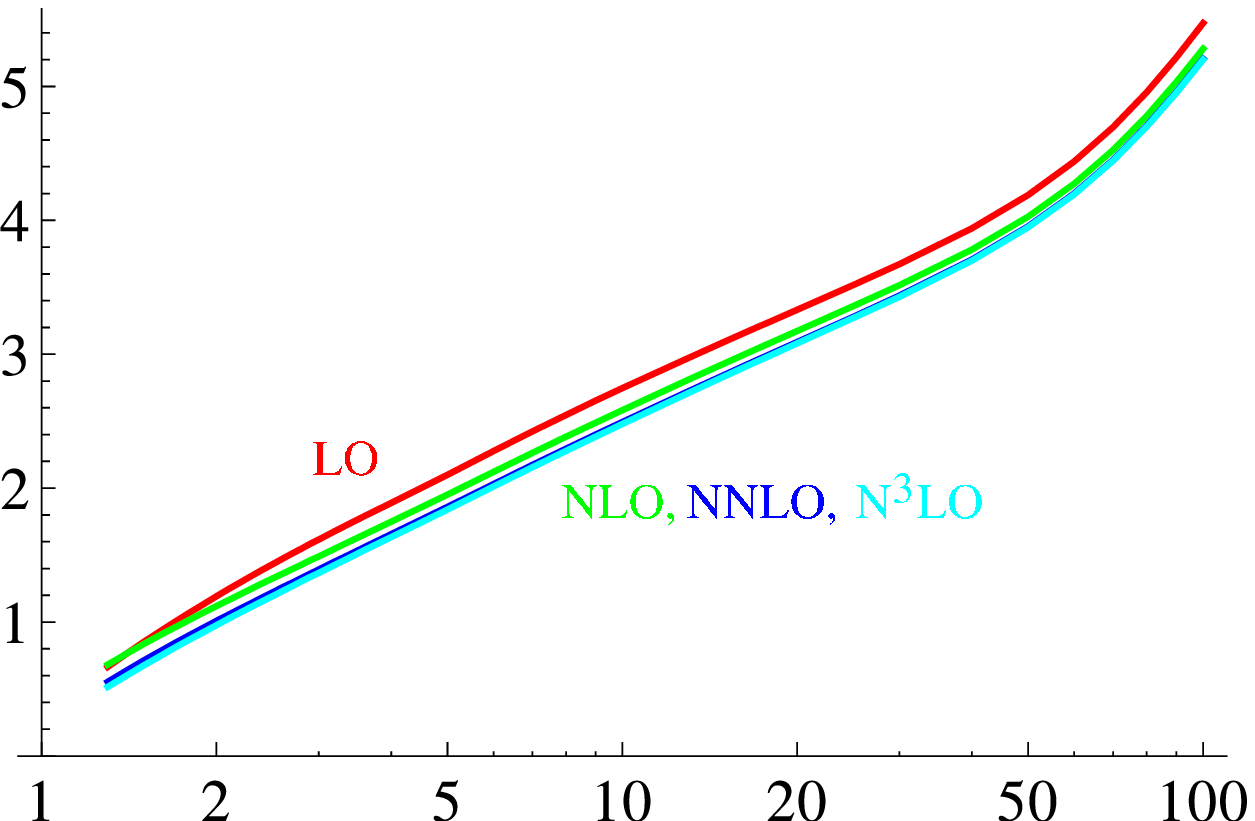}
\hfil
\includegraphics[width=0.30\textwidth]{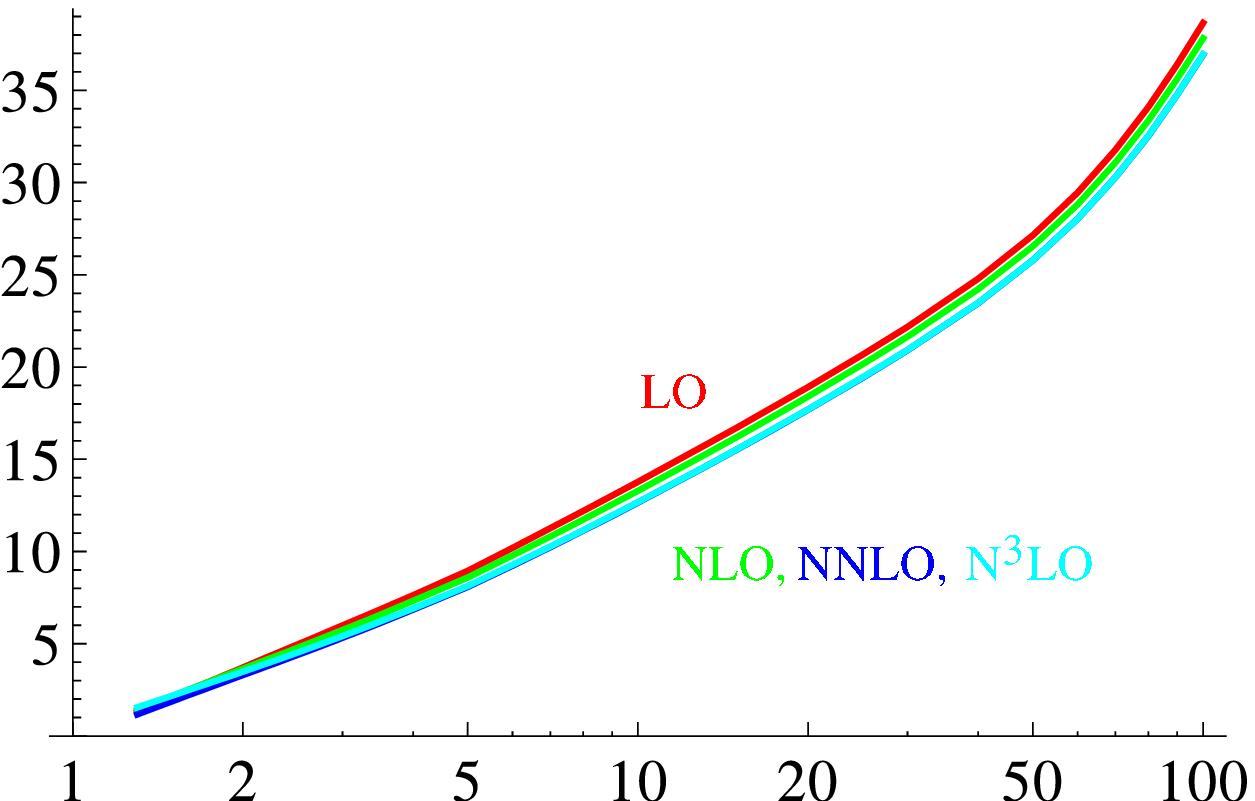}
\end{subfigure}
\begin{subfigure}{1.0\textwidth}
\caption{$F_{L}$ vs. $Q$.}
\label{fig:fLorders}
\includegraphics[width=0.30\textwidth]{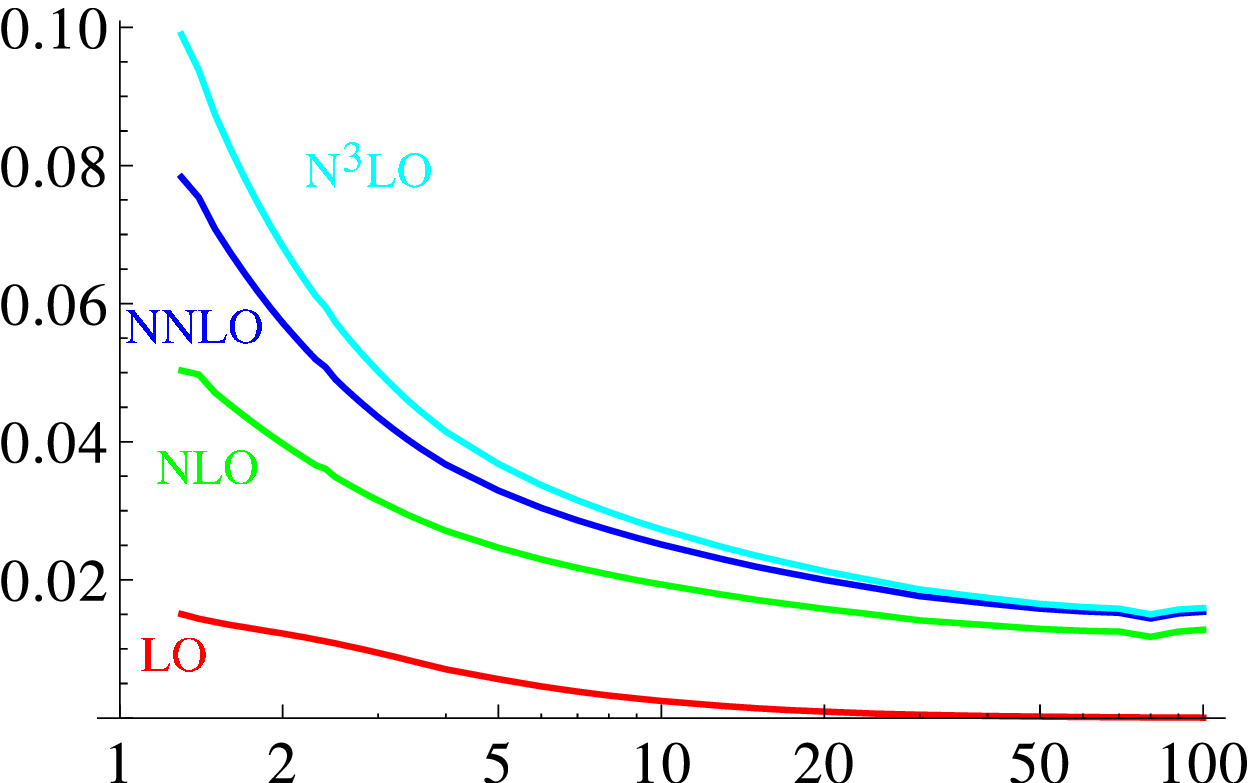}
\hfil
\includegraphics[width=0.30\textwidth]{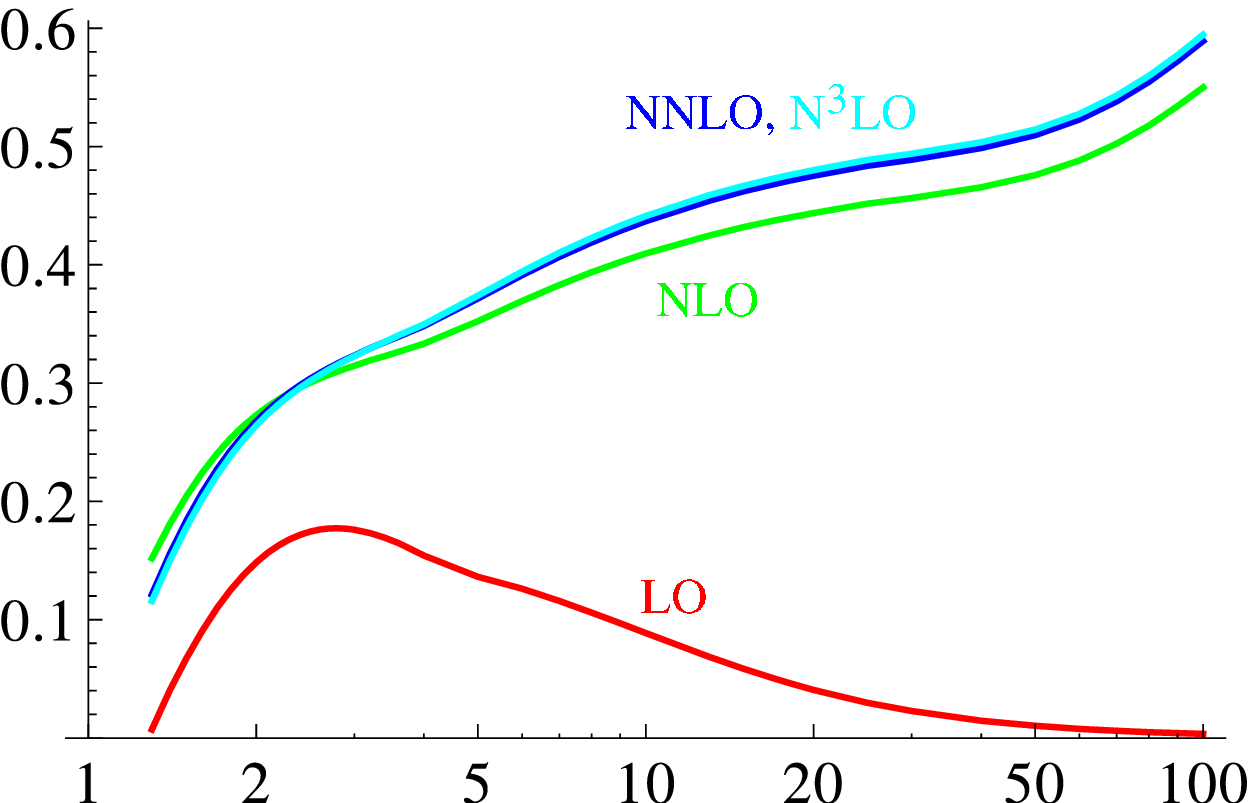}
\hfil
\includegraphics[width=0.30\textwidth]{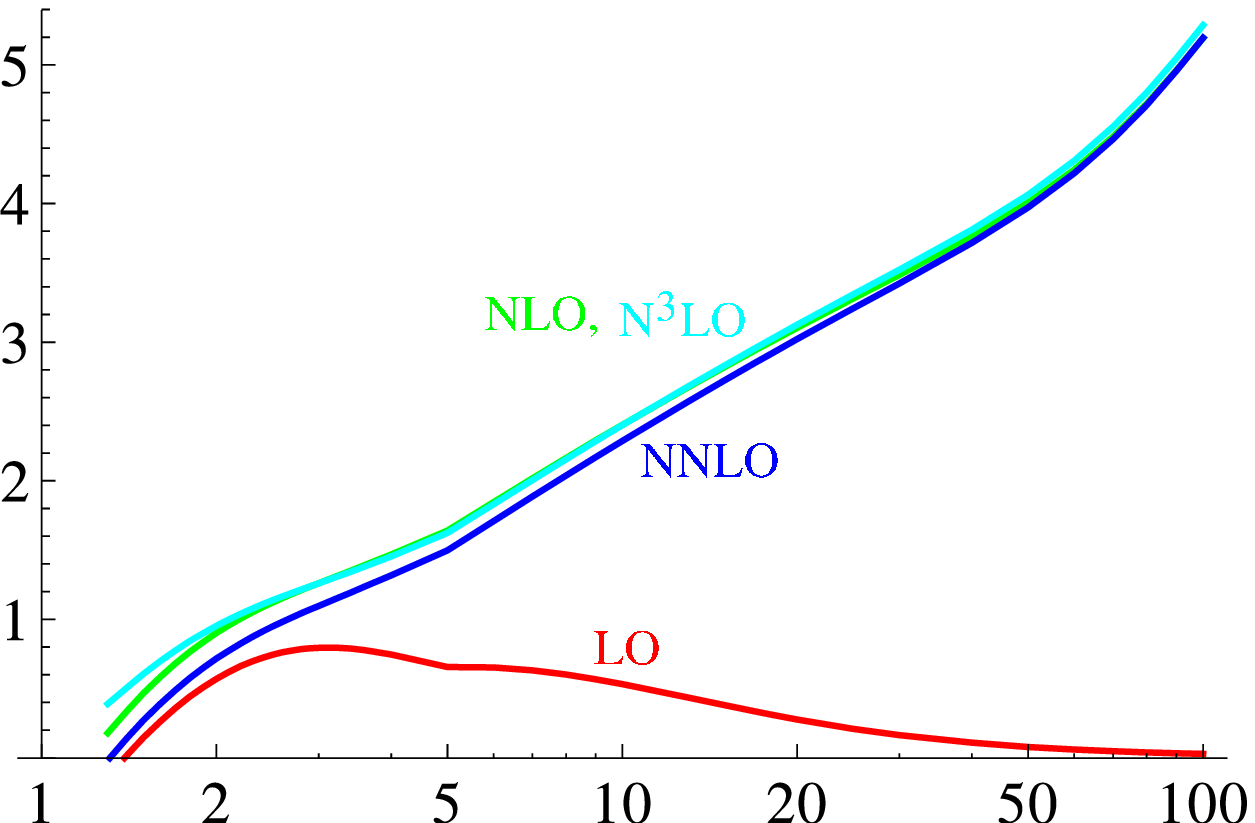}
\end{subfigure}
\caption{$F_{2,L}$ vs. $Q$ at \{LO, NLO, NNLO, N$^3$LO\}
(red, green, blue, cyan) for fixed $x=\{10^{-1},10^{-3},10^{-5}\}$
(left to right) for $n=2$ scaling.}
\end{figure*}

In Figure~\ref{fig:f2orders} we display the results for $F_{2}$
vs. $Q$ computed at various orders. For large $x$ (c.f. $x=0.1$) we
find the perturbative calculation is particularly stable; we see
that the LO result is within 20\% of the others at small $Q$, and
within 5\% at large $Q$. The NLO is within 2\% at small $Q$, and
indistinguishable from the NNLO and N$^3$LO for $Q$ values above $\sim10$~GeV.
The NNLO and N$^3$LO results are essentially identical throughout the
kinematic range. 

In Figure~\ref{fig:fLorders} we display the results for $F_{L}$
vs. $Q$ computed at various orders. In contrast to $F_{2}$,
we find the NLO corrections are large for $F_L$; this is because the LO
$F_{L}$ contribution (which violates the Callan-Gross relation) is suppressed
by $(m^{2}/Q^{2})$ compared to the dominant gluon contributions which
enter at NLO. Consequently, we observe (as expected) that the LO result
for $F_{L}$ receives large contributions from the higher order 
terms.
Essentially, the NLO is the first non-trivial order for $F_{L}$, and
the subsequent contributions then converge.
While the calculation of $F_L$ is certainly more challenging, 
the relevant
kinematic range probed by HERA the theoretical calculation is generally
stable.

\section{Conclusions\label{sec:conclusion} }

We extended the ACOT calculation for DIS structure functions 
to  N$^3$LO by combining  the exact ACOT
result at NLO with a $\chi(n)$-rescaling for the higher order terms; this allows us to
include the leading mass dependence at NNLO and N$^3$LO.

We studied the $F_2$ and $F_L$ structure functions as a function of
$x$ and $Q$.  We examined the flavor decomposition of these structure
functions, and verified that the heavy quarks were appropriately
suppressed in the low $Q$ region.
We found the results for $F_2$ were very stable across the full kinematic range
for $\{x,Q\}$, and the contributions from the NNLO and N${}^3$LO terms
were small.
For $F_L$, the higher order terms gave a proportionally larger
contribution (due to the suppression of the LO term from the
Callan-Gross relation); nevertheless, the contributions from the NNLO
and N${}^3$LO terms were generally small in the region probed by HERA.
Using the results of this calculation we can obtain  precise predictions for
the inclusive $F_2$ and $F_L$ structure functions which can be used
to analyze the HERA data.

\section*{Acknowledgments}

We thank 
M.~Botje, 
A.~M.~Cooper-Sarkar, 
A.~Glazov,
C.~Keppel, 
J.~G.~Morf\'in, 
P.~Nadolsky, 
M.~Guzzi,
J.~F.~Owens,
V.~A.~Radescu,
and
A.~Vogt
for discussions.
We 
acknowledge the hospitality of 
CERN, DESY, Fermilab, Les Houches, 
 Galileo Galilei Institute and the INFN.
This work was partially supported by the U.S.\ Department
of Energy under grant DE-FG02-04ER41299, the Lightner-Sams Foundation, 
the Th\'eorie LHC France initiative funded by the CNRS/IN2P3, and  
by  {\it Projet international de cooperation scientifique} PICS05854 between France and the USA.


{\raggedright
\begin{footnotesize}
 \bibliographystyle{DISproc}
 \bibliography{olness_fred.bib}
\end{footnotesize}
}


\end{document}